\documentclass[aps,prl,reprint,superscriptaddress,showpacs]{revtex4-1}
\usepackage{graphicx}
\usepackage{amsfonts,amssymb,amsmath}
\usepackage{color}
\usepackage{hyperref}

\begin{document}


\title{The electronic structure of the silicon vacancy color center in diamond}


\author{Christian Hepp}
\affiliation{Fachrichtung 7.2 (Experimentalphysik), Universit\"at des Saarlandes, Campus E2.6, 66123 Saarbr\"ucken, Germany}
\author{Tina M\"uller}
\affiliation{Atomic, Mesoscopic and Optical Physics Group, Cavendish Laboratory, University of Cambridge,
JJ Thomson Ave, Cambridge CB3 0HE, United Kingdom}
\author{Victor Waselowski}
\affiliation{Departmento de Fisica, Pontificia Universidad Catolica de Chile, Santiago 7820436, Chile}
\author{Jonas N. Becker}
\affiliation{Fachrichtung 7.2 (Experimentalphysik), Universit\"at des Saarlandes, Campus E2.6, 66123 Saarbr\"ucken, Germany}
\author{Benjamin Pingault}
\affiliation{Atomic, Mesoscopic and Optical Physics Group, Cavendish Laboratory, University of Cambridge,
JJ Thomson Ave, Cambridge CB3 0HE, United Kingdom}
\author{Hadwig Sternschulte}
\affiliation{KOMET RHOBEST GmbH, 6020 Innsbruck, Austria}
\affiliation{Fakult\"at f\"ur Physik, Technische Universit\"at M\"unchen, James-Franck-Strasse 1, 85748 Garching, Germany}
\author{Doris Steinm\"uller-Nethl}
\affiliation{KOMET RHOBEST GmbH, 6020 Innsbruck, Austria}
\author{Adam Gali}
\affiliation{Department of Atomic Physics, Budapest University of Technology and Economics, H-1111 Budapest, Hungary}
\affiliation{Institute for Solid State Physics and Optics, Wigner Research Centre for Physics, Hungarian Academy of Sciences, P.O. Box 49, H-1525 Budapest, Hungary}
\author{Jeronimo R. Maze}
\affiliation{Departmento de Fisica, Pontificia Universidad Catolica de Chile, Santiago 7820436, Chile}
\author{Mete Atat\"ure}
\affiliation{Atomic, Mesoscopic and Optical Physics Group, Cavendish Laboratory, University of Cambridge,
JJ Thomson Ave, Cambridge CB3 0HE, United Kingdom}
\author{Christoph~Becher}
\email[]{christoph.becher@physik.uni-saarland.de}
\affiliation{Fachrichtung 7.2 (Experimentalphysik), Universit\"at des Saarlandes, Campus E2.6, 66123 Saarbr\"ucken, Germany}



\newcommand{\bra}[1]{\left< #1 \right|}
\newcommand{\ket}[1]{\left| #1 \right>}
\newcommand{\braket}[2]{\left< #1 \right|\left. #2 \right>}
\newcommand{\matrixelement}[3]{\left< #1 | #2 | #3 \right>}

\date{\today}

\begin{abstract}
The negatively charged silicon vacancy (SiV) color center in diamond has recently proven its suitability for bright and stable single photon emission. However, its electronic structure so far has remained elusive. We here explore the electronic structure by exposing single SiV defects to a magnetic field where the Zeeman effect lifts the degeneracy of magnetic sublevels. The similar response of  single centers and a SiV ensemble in a low strain reference sample proves our ability to fabricate almost perfect single SiVs, revealing the true nature of the defect's electronic properties. We model the electronic states using a group-theoretical approach yielding a good agreement with the experimental observations. Furthermore, the model correctly predicts polarization measurements on single SiV centers and explains recently discovered spin selective excitation of SiV defects.
\end{abstract}


\maketitle


Negatively charged silicon vacancy (SiV$^{-}$) color centers in diamond show a typical room-temperature zero phonon line (ZPL) at 738 nm which splits into a four line fine structure centered at about 737 nm when cooled down to liquid helium temperature \cite{Clark1995, Sternschulte1995, Neu2013a}. The origin of the fine structure splitting is attributed to a split ground and excited state \cite{Clark1995}. One mechanism that can account for the level splitting is spin-orbit (SO) coupling, like it is present for the excited state in negatively charged nitrogen-vacancy ($\text{NV}^-$) centers \cite{Maze2011}. Alternatively, Clark et al.\ \cite{Clark1995} and Moliver \cite{Moliver2003} suggest a tunnel splitting whereas Goss et al.\ \cite{Goss1996} assume a Jahn-Teller (JT) effect in addition to SO coupling to lift the orbital degeneracy between the electronic states which account for the presumed optical transition $^2E_u \to {^2}E_g$.  To form doubly degenerate $^2E$ many-body wave functions, at least a trigonal defect symmetry is required \cite{Tinkham1964,Walker1979}.
The molecular structure of the SiV center was predicted using density functional theory (DFT) to show a rather unique split vacancy configuration, exhibiting a $D_{\text{3d}}$ symmetry \cite{Goss2007}. Yet, polarization \cite{Brown1995,Neu2011b} and uniaxial stress measurements \cite{Sternschulte1995} evidenced lower symmetrical point groups such as $C_2$ or $D_2$ symmetry. Still, all these experimental evidences were obtained using samples that possess strongly strained environments for the defect centers. In this letter, however, we present evidence for the predicted $D_{\text{3d}}$ symmetry by performing spectroscopy on SiV centers in low-strain samples.

Recently published EPR measurements showed that the presumed neutral charge state $\text{SiV}^0$ is a $S = 1$ system \cite{DHaenens2011}. This suggests that its negative counterpart $\text{SiV}^-$ is a paramagnetic $S = 1/2$ system, although this has not been confirmed by independent EPR measurements so far. Very recently, we reported direct spin-selective population of the SiV$^{-}$ excited states under a magnetic field, resulting in a spin-tagged resonance fluorescence pattern \cite{Mueller2013}, suggesting that the SiV$^{-}$ shows effectively $S = 1/2$. In the present letter, we experimentally explore the electronic states of the SiV center by measuring Zeeman splittings and polarization orientation of the fine structure lines. A detailed theoretical analysis of the SiV$^{-}$ center allows for an assignment of electronic states and correctly describes the Zeeman splittings, polarization properties and earlier measurements on spin-selective excitation \cite{Mueller2013}.

Crucial prerequisites for the experimental investigation of the SiV electronic states are the availability of SiV centers in low strain samples (referred to as ``ideal'' centers) and the ability to observe isolated single centers in order to prevent inhomogeneous broadening effects. Two samples were investigated: The first (``SiV ensemble sample'') is a thin single crystalline diamond film which contains a large ensemble of SiV defects. The second sample (``SIL sample'') is a high purity bulk diamond in which single SiV centers were created using ion implantation. To enhance the collection of the single emitter fluorescence, an array of solid immersion lenses (SIL) was fabricated using focussed ion beam milling (for details about the samples see Supplemental Material \cite{Supplementary}). The samples were investigated in two homebuilt confocal microscopes, with excitation wavelengths in the range 690 - 700nm, one equipped with a superconducting magnet mounted in Faraday configuration, providing fields up to 7 T. Throughout all measurements, the magnetic field is aligned parallel to the crystallographic $[001]$ direction.

\begin{figure}[hb]
\centering
\includegraphics[scale=0.5]{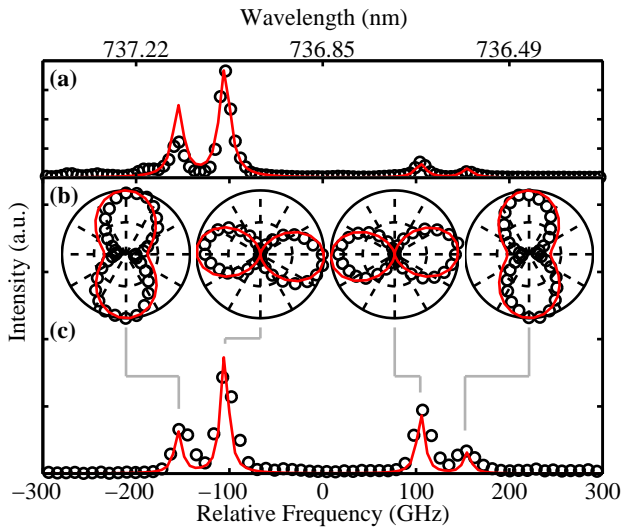}
\caption{(Color online) Spectral fine structure of (a) an ensemble of SiV centers at 4 K and (c) single emitter SIL1 under a solid immersion lens at 18 K. The inset (b) shows the polarization of each single fine structure line. Black dots: measurements, red solid lines: simulations. }
\label{fig:SpectralFineStructure}
\end{figure}

Figure \ref{fig:SpectralFineStructure} shows the zero-field spectral fine structure of an ensemble of SiV$^{-}$ defects (Fig.~\ref{fig:SpectralFineStructure}(a)) and a single defect under a SIL (SIL1, Fig.~\ref{fig:SpectralFineStructure}(c)). The position and splitting (ground state splitting: $\Delta E_\text{g} = 50$ GHz (0.21 meV), excited state splitting: $\Delta E_\text{e} = 260$ GHz (1.08 meV)) of the fine structure in the SiV$^{-}$ ensemble sample are in excellent agreement with former findings \cite{Clark1995}. The linewidth of the fine structure lines in the SiV$^{-}$ ensemble is $\approx 10$ GHz \cite{Neu2013a,Mueller2013}, indicating a very small inhomogeneous broadening and proving the high crystalline quality of the diamond film. Therefore, we treat this SiV$^{-}$ ensemble as the reference which we compare single SiV$^{-}$ centers to. For SIL1, the splitting of the two doublets is identical with the reference sample within our resolution limit of 5 GHz. The relative intensity of  the peaks is different from the SiV$^{-}$ ensemble which is due to a different temperature and resulting different thermalization \cite{Clark1995}.
%

To investigate the dipole transitions of the SiV$^{-}$, we measure the photoluminescence polarization for several single emitters in the SIL sample (e.g. SIL1, Fig.~\ref{fig:SpectralFineStructure}(b)). The polarization of the fine structure lines can be grouped in two subsets. The inner transitions are polarized parallel to each other and perpendicular to the outer ones, where all polarization axes appear along the $\left<110\right>$ crystal axis. For our measurements, corresponding to a projection into the $(001)$ plane, the observed polarization direction is consistent with the predicted $\left<111\right>$ orientation of the SiV in D$_{\text{3d}}$ symmetry \cite{Goss1996}. This result is confirmed by independent measurements on a larger number of SiV$^{-}$ defects \cite{Rogers2013}. The red solid lines in Fig.~\ref{fig:SpectralFineStructure}(c) represent a simulation using the model developed below.

\begin{figure*}
\centering
\includegraphics[scale=0.5]{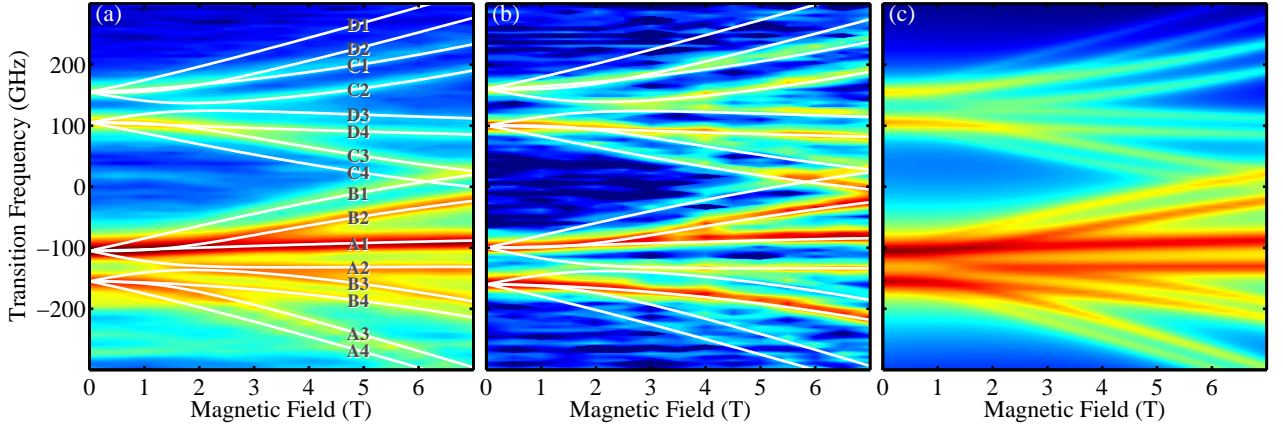}
\caption{(Color online) Spectral fine structure splitting of (a) an SiV$^{-}$ ensemble (contour plot, color coding indicates peak intensity in logarithmic a.u.) and (b) single SiV$^{-}$ defect under a SIL (SIL2) vs. applied magnetic field (in $(001)$ direction). White solid lines are calculated transitions based on the model mentioned in the text. The labeling of transitions is in accordance with Fig. \ref{fig:LevelSplitting}. Panel (c) displays a simulation of the fine structure lines intensity assuming dipolar transitions.}
\label{fig:MagneticFieldSplittings}
\end{figure*}
To gain further insight into the electronic structure, both the ensemble of SiV$^{-}$ and another single emitter, SIL2, were exposed to a magnetic field. The Zeeman effect leads to a splitting  of each fine structure line into four lines, where this splitting is not symmetrical and shows several avoided crossings (Fig.~\ref{fig:MagneticFieldSplittings}). The splitting into four components points towards a spin $1/2$ system, and the avoided crossings indicate SO coupling. From the fact that the ensemble spectrum shows a very similar splitting pattern to the single defect and splits into as many lines, we learn that all possible equivalent orientations of the center in the ensemble have the same relative angle to the magnetic field. The only orientation which allows for this fact is indeed a $\left<111\right>$-orientation - which shows for the first time experimental evidence of a $\left<111\right>$-orientation for SiV centers and establishes a link to theories published so far \cite{Goss1996,Goss2007}.

\begin{figure}
\centering
\includegraphics[scale=0.5]{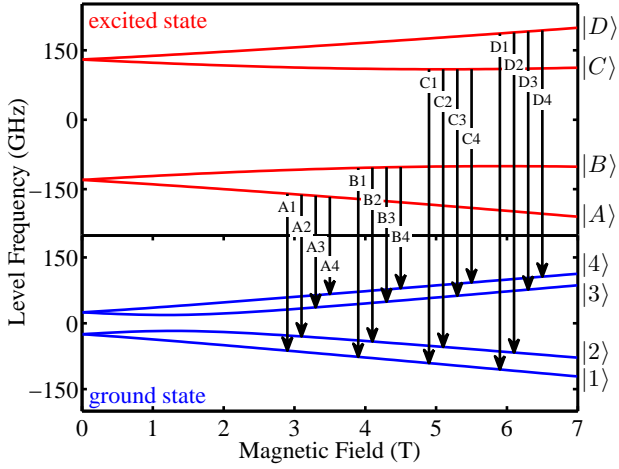}
\caption{(Color online) Calculated splitting of electronic levels with increasing magnetic field. Ground and excited state labelled according to the letters and numbers at the right of the panel. Optical transitions between all levels indicated by black arrows and correspond to white solid lines in Fig.~\ref{fig:MagneticFieldSplittings}(a).}
\label{fig:LevelSplitting}
\end{figure}

In the following section, we develop a model of the SiV center electronic structure. Starting from the experimental evidence, we model the SiV oriented along $\left<111\right>$ direction and assume $D_{\text{3d}}$ symmetry \cite{Goss1996,Goss2007}. The carbon dangling bond orbitals are superimposed to construct symmetry adapted linear combinations (SALCs) which form the electronic states of the SiV center and transform as the irreducible representations $A_\text{1g}$, $A_\text{2u}$, $E_\text{u}$ and $E_\text{g}$. The orbitals belonging to the Si atom can be approximated as hydrogen-like wave functions. DFT calculations yield the ordering of the dangling bond SALCs and Si states, respectively, indicating that the SALCs are considerably lower in energy \cite{Gali2013}, and that only these need to be considered for optically active transitions \cite{DHaenens2011}. The center hosts a total number of eleven electrons: six electrons contributed by dangling bonds, four electrons from the Si-atom and an one electron trapped from nearby donors to account for the negative charge \cite{DHaenens2011}. Taking into account spin degeneracy, the $A$ $(E)$ states accommodate 2 (4) electrons, i.e. one unpaired electron remains in the $E_\text{g}$ state. We therefore consider the SiV$^{-}$ ground state as $^2E_\text{g}$ and the excited state as $^2E_\text{u}$ where a single hole formalism is equivalent to a single electron state except for signs in spin-orbit interaction \cite{Maze2011}. The $^2E$-states have a twofold orbital and a twofold spin degeneracy. They can be split into (purely spin degenerate) states by either SO or JT interaction. This results in the split ground and excited states with 4 possible optical transitions forming the zero-field ZPL fine structure (Fig.~\ref{fig:SpectralFineStructure}). In a magnetic field Zeeman interaction terms $\mathbb{H}^\text{Z,L}_\text{g,e}$ and $\mathbb{H}^\text{Z,S}_\text{g,e}$ both lift the spin degeneracy ($\mathbb{H}^\text{Z,L}_\text{g,e}$ in conjunction with SO coupling). SO, JT and Zeeman interaction sum up to the following total Hamiltonian for ground or excited state:
\begin{eqnarray}
\mathbb{H_\text{g,e}} &=& \mathbb{H}_\text{g,e}^0 + \mathbb{H}^{\text{SO}}_{\text{g,e}} + \mathbb{H}^\text{JT}_\text{g,e} + \mathbb{H}^\text{Z,L}_\text{g,e} + \mathbb{H}^\text{Z,S}_\text{g,e} \label{eqn:Hamiltonian} \\
 &=& \mathbb{H}_\text{g,e}^0 + \lambda_\text{g,e} L_\text{z} S_\text{z} + \Upsilon_{\text{JT}} + f \gamma_\text{L}  L_\text{z} B_\text{z} + \gamma_\text{S} \vec S \cdot\vec B, \nonumber
\end{eqnarray}

\begin{figure*}[ht]
\centering
\includegraphics[scale=0.5]{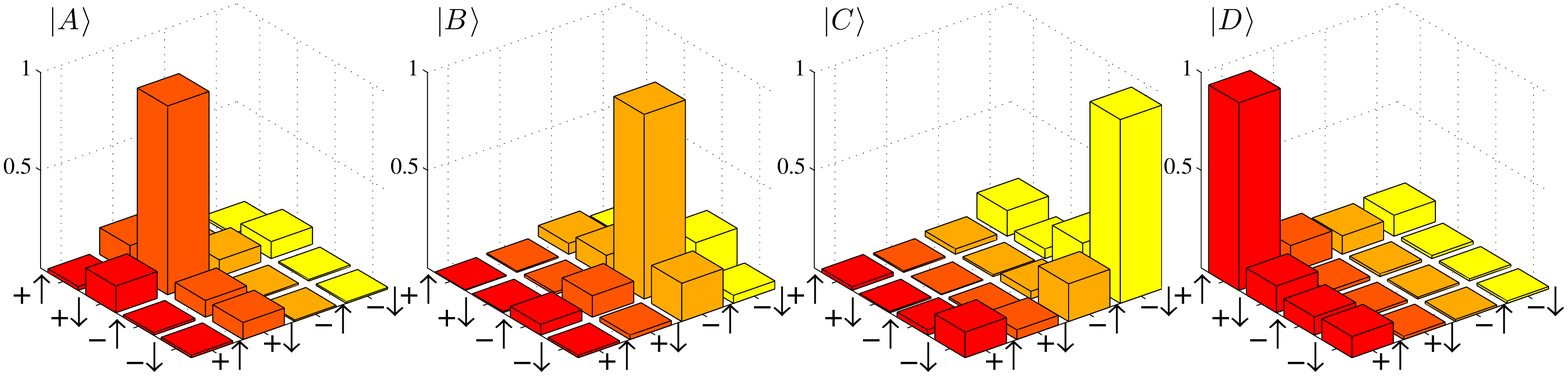}
\caption{(Color online) Tomography of the excited states at $B=4$ T using the model parameters, which lead to the level splitting depicted in Fig.~\ref{fig:LevelSplitting}. Basis states given in the eigenstates of $L_\text{z}$ operator $ \left| e_\pm \right> = -(\left| e_\text{x} \right> \mp i\left| e_\text{y} \right>$ and $S_\text{z}$ operator $\left|\uparrow \right> ,\,\left|\downarrow \right>$.}
\label{fig:StateTomography}
\end{figure*}
where $\mathbb{H}_\text{g,e}^0 $ is the non-perturbed Hamiltonian, $\lambda_\text{g,e}$ is the SO coupling constant and $\gamma_\text{L} = \mu_B/\hbar$, $\gamma_\text{S}=2\mu_B/\hbar$ are orbital and electron gyromagnetic ratios. We express all matrices in the $\left\{ \ket{ e_\text{gx} \uparrow }, \ket{ e_\text{gx} \downarrow }, \ket{ e_\text{gy} \uparrow }, \ket{ e_\text{gy} \downarrow } \right\}$ basis for the ground state and $\left\{ \ket{ e_\text{ux} \uparrow }, \ket{ e_\text{ux} \downarrow }, \ket{ e_\text{uy} \uparrow }, \ket{ e_\text{uy} \downarrow } \right\}$ for the excited state: Orbital operators $\matrixelement{e}{L_{\text{r}}}{e}$, $r = x,y,z$, can be directly deduced from the group theoretical description of the NV$^{-}$ color center in diamond \cite{Maze2011}.
$\vec{S} = (\sigma_{x},\sigma_{y},\sigma_{z})$ are  Pauli spin matrices and $\Upsilon_{\text{JT}}$ denotes a  $E \otimes e$ linear vibronic JT coupling \cite{Abtew2011, Rorison1984, LonguetHiggins1958}, where we define the JT coupling strength $\Upsilon_\text{g,e} =(\Upsilon_{x}^{2} + \Upsilon_{y}^{2} )^{1/2}$ for ground and excited state respectively. We tentatively suggest a factor $f$ which accounts for quenching of the orbital gyromagnetic factor due to JT interaction - a common effect for solid state defect systems \cite{Ham1965,Tosatti1996}. The coordinate frame (internal reference frame) is given by the high symmetry $\left<111\right>$ axis of the SiV which we denote as $z$ axis and $x$, $y$ in the $(111)$ plane. We have reduced the SO and Zeeman coupling to the expressions in Eq.~\eqref{eqn:Hamiltonian} because under D$_{\text{3d}}$ symmetry, $L_{\text{x}}$ and $L_{\text{y}}$ only affect states which transform as $A_{\text{1g}}$ and $A_{\text{2u}}$. For each electronic configuration (ground and excited state), the splitting is given by both the SO and JT interactions and is equal to $\left(\lambda_\text{g,e}^2 + 4\Upsilon_\text{g,e}^2\right)^{1/2}$. This quantity is set equal to the experimentally observed ground and excited state splitting and we use $r_\text{g,e} = \lambda_\text{g,e} / \Upsilon_\text{g,e}$ as free parameters \cite{Supplementary}.

Solving the secular equation defined by the Hamiltonian in Eq.~\eqref{eqn:Hamiltonian} yields the energies of each state at a given magnetic field value (i.e.\ the eigenvalues, Fig.~\ref{fig:LevelSplitting}) as well as their eigenvectors $\ket{1}$-$\ket{4}$ for ground state and $\ket{A}$ -$\ket{D}$ for excited state, respectively. We calculate the optical transition frequencies between the electronic levels (arrows in Fig.~\ref{fig:LevelSplitting}, white solid lines in Fig.~\ref{fig:MagneticFieldSplittings}(a)) and compare them with the Zeeman spectrum of our reference sample (Fig.~\ref{fig:MagneticFieldSplittings}(a)). Varying only the quenching factor $f$ and the ratio $r_g$, $r_e$ between SO and JT in ground and excited state, respectively, we iteratively fit transition frequencies to the experimental data \cite{Supplementary}. In both ground and excited state we observe a high SO coupling and only little JT distortion ($r_g= 3.9$, $r_e = 6.2$). As a consequence of the strong SO coupling, Fig.~\ref{fig:LevelSplitting} shows a avoided level crossing at 2 T in the ground state which is also a predominant feature in the magnetic field splitting spectra (Fig.~\ref{fig:MagneticFieldSplittings}(a,b)). We note that the fit parameters for the reference sample (Tab.~\ref{tab:FitParameters}) are identical with the parameters for single center SIL2 (within our measurement resolution). This proves our ability to deterministically fabricate ``ideal'' single SiV centers which reproduce many experimental \cite{Sternschulte1995,Clark1995} and theoretical \cite{Goss2007} results.

The theoretical model derived here can also be used to assess the applicability of the SiV$^{-}$ center as a quantum bit, i.e.\ the accessibility of a well defined spin state, as indicated by recent experiments on spin-selective population of SiV$^{-}$ excited states \cite{Mueller2013}. We are now able to elucidate this phenomenon by calculating the spin and orbital eigenstates of the excited state  $\ket{A}$ -$\ket{D}$. Figure \ref{fig:StateTomography}  displays the absolute value of the excited state density matrix, expressed in basis states of the $L_\text{z}$ and $S_\text{z}$ operators. As the $[001]$ crystal axis is aligned along the magnetic field direction, the angle between magnetic field and high symmetry axis of the SiV defect is fixed to $54.7^{\circ}$. Therefore, one would expect that Zeeman interaction with off-axis magnetic field terms $B_\text{x}$, $B_\text{y}$ leads to spin mixing both in ground and excited state. It is noticeable, however, that in the excited state all electronic states are dominated by contributions from one spin state with a population probability of above 97\%, respectively (Fig.~\ref{fig:StateTomography}). The reason for this particular spin polarization is the strong quenching $f$ of the orbital magnetic momentum. In consequence, the spin effectively remains a good quantum number. In contrast, all four ground states show stronger mixing with maximum spin polarization of 80\% \cite{Supplementary}. We note that the resulting spin orientations in the presence of magnetic field are in full agreement with those reported for spin selective resonant excitation \cite{Mueller2013}. As the reason for the spin mixing is given by off-axis magnetic field terms, the spin state purity can be further increased by aligning the magnetic field along the SiV high symmetry axis $\left<111\right>$ yielding predicted ground state spin polarizations of above 91\% \cite{Supplementary}.

 \begin{table}
 \caption{Fit parameters for SiV reference ensemble (``ensemble'' sample) and single defects depicted above.\label{tab:FitParameters}}
 \begin{ruledtabular}
 \begin{tabular}{cccccccccc}
 Emitter & $f$ & $\lambda_\text{g}$ & $\Upsilon_\text{x,g}$ &  $\Upsilon_\text{y,g}$ & $\lambda_\text{e}$ & $\Upsilon_\text{x,e}$ & $\Upsilon_\text{y,e}$ & $\Delta E_g$ & $\Delta E_e$\\
 name& (-) & \multicolumn{8}{c}{(GHz)}\\
 \hline
 Ensemble 	& 		0.1	&	 45 	&	 \multicolumn{2}{c}{11} &		257	&		\multicolumn{2}{c}{20} 	&		50		&		260	\\
 SIL1  (Fig. \ref{fig:SpectralFineStructure}(c)) &	0.1		   &		49	&		 2	&	3	&		257		&		12	&  16 & 50		& 260 \\
 SIL2 (Fig. \ref{fig:MagneticFieldSplittings}(b)) &	0.1		   &		54	&	 \multicolumn{2}{c}{14} &			257		&		\multicolumn{2}{c}{20}   & 60		& 260
\end{tabular}
\end{ruledtabular}
\end{table}
To further verify the proposed theoretical model, we infer the change in angular momentum for the observed dipole transitions  and calculate the expected polarization of the emitted light. The dipole moment $\mathbf{d} = \matrixelement{f}{\mathbf{r}}{i} $, with $\ket{i} = \ket{A} \ldots \ket{D} $ and $\bra{f} = \bra{1} \ldots \bra{4} $, is directly derived from the numerically determined eigenvectors. We then project the emitted linear and circular polarization components onto our observation plane $(001)$ and use the numerical method of Ref.~\cite{Neu2012b} to estimate the relative collection efficiencies of our experimental setup for different dipole components. The result includes both the polarization direction and visibility (red line in Fig.~\ref{fig:SpectralFineStructure}). Starting from parameters of an ideal SiV$^{-}$ (Tab.~\ref{tab:FitParameters}, ``ensemble''), we fit the fluorescence polarization of emitter SIL1 (Fig.~\ref{fig:SpectralFineStructure}(c)) and obtain comparable parameters (Tab.~\ref{tab:FitParameters}, ``SIL1''). We note, that the polarization of the first and second peak is tilted by $8\pm4^{\circ}$ away from $\left<110\right>$ direction. We model this polarization change by adjusting JT distortion parameters $\Upsilon_{x}$ and $\Upsilon_{y}$; a similar result would be obtained using a static strain addition.  The calculation of the dipole transition strength further allows the reconstruction of the Zeeman spectrum at arbitrary magnetic field values (Fig.~\ref{fig:MagneticFieldSplittings}(c)) yielding an impressive agreement with the measured data (Fig.~\ref{fig:MagneticFieldSplittings}(a)).

We note that the spectra and polarization graphs of Fig.~\ref{fig:SpectralFineStructure}(a-c) correspond to the case of ideal, strain-free SiV centers. Different Zeeman splitting patterns as well as polarization directions and visibilities have been experimentally observed in other regions of the SIL sample and in nanodiamonds. These deviations can be explained by the influence of crystal strain in the host lattice. The addition of a phenomenological strain Hamiltonian to the theoretical model faithfully reproduces these variations (to be published elsewhere). The variation of polarization direction under strain further might explain the lower defect symmetry inferred from SiV$^{-}$ polarization data in strained diamond samples \cite{Brown1995,Neu2011b,Sternschulte1995}.

In conclusion, we demonstrated the fabrication of unstrained, single SiV$^{-}$ centers that allow for the study of the center's unperturbed electronic structure. Our theoretical model explains both the SiV$^{-}$ fine structure splitting in magnetic fields as well as the polarization of the zero field fine structure components. Furthermore, it provides a qualitative explanation of the first spin-related experiments of the negatively charged SiV. This profound understanding paves the way for utilizing the SiV$^{-}$ center in quantum information applications.

\begin{acknowledgments}
We thank David Steinmetz and Jan Meijer for ion implantation of the SILs sample and Elke Neu for helpful discussions. Ion implantation was performed at and supported by RUBION, central unit of the Ruhr-Universit\"at Bochum. We are deeply grateful to Christoph Pauly and Frank M\"ucklich from the Functional Materials Group at the Saarland University for SIL fabrication and Annika S. Diehl for automizing polarization measurements. C.B. acknowledges funding from the Bundesministerium f\"ur Bildung und Forschung within the projects EPHQUAM (Contract No. 01BL0903) and QuOReP (Contract No. 01BQ1011). M.A. gratefully acknowledges financial support by the University of Cambridge, the European Research Council (FP7/2007-2013)/ERC Grant agreement no. 209636 and FP7 Marie Curie Initial Training Network S$^{3}$NANO. A.G. acknowledges the support from EU Commission by the FP7 DIAMANT project (No. 270197). V.W. thanks CONICYT-PCHA/Doctorado Nacional/2013-21130747 for financial support. J.R.M. acknowledges support from Concicyt Fondecyt Iniciaci—n no. 11100265 and PIA no. ACT1108..
\end{acknowledgments}


\begin{thebibliography}{22}%
\makeatletter
\providecommand \@ifxundefined [1]{%
 \@ifx{#1\undefined}
}%
\providecommand \@ifnum [1]{%
 \ifnum #1\expandafter \@firstoftwo
 \else \expandafter \@secondoftwo
 \fi
}%
\providecommand \@ifx [1]{%
 \ifx #1\expandafter \@firstoftwo
 \else \expandafter \@secondoftwo
 \fi
}%
\providecommand \natexlab [1]{#1}%
\providecommand \enquote  [1]{``#1''}%
\providecommand \bibnamefont  [1]{#1}%
\providecommand \bibfnamefont [1]{#1}%
\providecommand \citenamefont [1]{#1}%
\providecommand \href@noop [0]{\@secondoftwo}%
\providecommand \href [0]{\begingroup \@sanitize@url \@href}%
\providecommand \@href[1]{\@@startlink{#1}\@@href}%
\providecommand \@@href[1]{\endgroup#1\@@endlink}%
\providecommand \@sanitize@url [0]{\catcode `\\12\catcode `\$12\catcode
  `\&12\catcode `\#12\catcode `\^12\catcode `\_12\catcode `\%12\relax}%
\providecommand \@@startlink[1]{}%
\providecommand \@@endlink[0]{}%
\providecommand \url  [0]{\begingroup\@sanitize@url \@url }%
\providecommand \@url [1]{\endgroup\@href {#1}{\urlprefix }}%
\providecommand \urlprefix  [0]{URL }%
\providecommand \Eprint [0]{\href }%
\providecommand \doibase [0]{http://dx.doi.org/}%
\providecommand \selectlanguage [0]{\@gobble}%
\providecommand \bibinfo  [0]{\@secondoftwo}%
\providecommand \bibfield  [0]{\@secondoftwo}%
\providecommand \translation [1]{[#1]}%
\providecommand \BibitemOpen [0]{}%
\providecommand \bibitemStop [0]{}%
\providecommand \bibitemNoStop [0]{.\EOS\space}%
\providecommand \EOS [0]{\spacefactor3000\relax}%
\providecommand \BibitemShut  [1]{\csname bibitem#1\endcsname}%
\let\auto@bib@innerbib\@empty
\bibitem [{\citenamefont {Clark}\ \emph {et~al.}(1995)\citenamefont {Clark},
  \citenamefont {Kanda}, \citenamefont {Kiflawi},\ and\ \citenamefont
  {Sittas}}]{Clark1995}%
  \BibitemOpen
  \bibfield  {author} {\bibinfo {author} {\bibfnamefont {C.~D.}\ \bibnamefont
  {Clark}}, \bibinfo {author} {\bibfnamefont {H.}~\bibnamefont {Kanda}},
  \bibinfo {author} {\bibfnamefont {I.}~\bibnamefont {Kiflawi}}, \ and\
  \bibinfo {author} {\bibfnamefont {G.}~\bibnamefont {Sittas}},\ }\href@noop {}
  {\bibfield  {journal} {\bibinfo  {journal} {Phys. Rev. B}\ }\textbf {\bibinfo
  {volume} {51}},\ \bibinfo {pages} {16681} (\bibinfo {year}
  {1995})}\BibitemShut {NoStop}%
\bibitem [{\citenamefont {Sternschulte}\ \emph {et~al.}(1995)\citenamefont
  {Sternschulte}, \citenamefont {Thonke}, \citenamefont {Gerster},
  \citenamefont {Limmer}, \citenamefont {Sauer}, \citenamefont {Spitzer},\ and\
  \citenamefont {M\"{u}nzinger}}]{Sternschulte1995}%
  \BibitemOpen
  \bibfield  {author} {\bibinfo {author} {\bibfnamefont {H.}~\bibnamefont
  {Sternschulte}}, \bibinfo {author} {\bibfnamefont {K.}~\bibnamefont
  {Thonke}}, \bibinfo {author} {\bibfnamefont {J.}~\bibnamefont {Gerster}},
  \bibinfo {author} {\bibfnamefont {W.}~\bibnamefont {Limmer}}, \bibinfo
  {author} {\bibfnamefont {R.}~\bibnamefont {Sauer}}, \bibinfo {author}
  {\bibfnamefont {J.}~\bibnamefont {Spitzer}}, \ and\ \bibinfo {author}
  {\bibfnamefont {P.~C.}\ \bibnamefont {M\"{u}nzinger}},\ }\href {\doibase
  http://dx.doi.org/10.1016/0925-9635(95)00298-7} {\bibfield  {journal}
  {\bibinfo  {journal} {Diam. Relat. Mater.}\ }\textbf {\bibinfo {volume}
  {4}},\ \bibinfo {pages} {1189} (\bibinfo {year} {1995})}\BibitemShut
  {NoStop}%
\bibitem [{\citenamefont {Neu}\ \emph {et~al.}(2013)\citenamefont {Neu},
  \citenamefont {Hepp}, \citenamefont {Hauschild}, \citenamefont {Gsell},
  \citenamefont {Fischer}, \citenamefont {Sternschulte}, \citenamefont
  {Steinm\"{u}ller-Nethl}, \citenamefont {Schreck},\ and\ \citenamefont
  {Becher}}]{Neu2013a}%
  \BibitemOpen
  \bibfield  {author} {\bibinfo {author} {\bibfnamefont {E.}~\bibnamefont
  {Neu}}, \bibinfo {author} {\bibfnamefont {C.}~\bibnamefont {Hepp}}, \bibinfo
  {author} {\bibfnamefont {M.}~\bibnamefont {Hauschild}}, \bibinfo {author}
  {\bibfnamefont {S.}~\bibnamefont {Gsell}}, \bibinfo {author} {\bibfnamefont
  {M.}~\bibnamefont {Fischer}}, \bibinfo {author} {\bibfnamefont
  {H.}~\bibnamefont {Sternschulte}}, \bibinfo {author} {\bibfnamefont
  {D.}~\bibnamefont {Steinm\"{u}ller-Nethl}}, \bibinfo {author} {\bibfnamefont
  {M.}~\bibnamefont {Schreck}}, \ and\ \bibinfo {author} {\bibfnamefont
  {C.}~\bibnamefont {Becher}},\ }\href {\doibase 10.1088/1367-2630/15/4/043005}
  {\bibfield  {journal} {\bibinfo  {journal} {New J. Phys.}\ }\textbf {\bibinfo
  {volume} {15}},\ \bibinfo {pages} {043005} (\bibinfo {year}
  {2013})}\BibitemShut {NoStop}%
\bibitem [{\citenamefont {Maze}\ \emph {et~al.}(2011)\citenamefont {Maze},
  \citenamefont {Gali}, \citenamefont {Togan}, \citenamefont {Chu},
  \citenamefont {Trifonov}, \citenamefont {Kaxiras},\ and\ \citenamefont
  {Lukin}}]{Maze2011}%
  \BibitemOpen
  \bibfield  {author} {\bibinfo {author} {\bibfnamefont {J.~R.}\ \bibnamefont
  {Maze}}, \bibinfo {author} {\bibfnamefont {A.}~\bibnamefont {Gali}}, \bibinfo
  {author} {\bibfnamefont {E.}~\bibnamefont {Togan}}, \bibinfo {author}
  {\bibfnamefont {Y.}~\bibnamefont {Chu}}, \bibinfo {author} {\bibfnamefont
  {A.}~\bibnamefont {Trifonov}}, \bibinfo {author} {\bibfnamefont
  {E.}~\bibnamefont {Kaxiras}}, \ and\ \bibinfo {author} {\bibfnamefont
  {M.~D.}\ \bibnamefont {Lukin}},\ }\href {\doibase
  10.1088/1367-2630/13/2/025025} {\bibfield  {journal} {\bibinfo  {journal}
  {New J. Phys.}\ }\textbf {\bibinfo {volume} {13}},\ \bibinfo {pages} {025025}
  (\bibinfo {year} {2011})}\BibitemShut {NoStop}%
\bibitem [{\citenamefont {Moliver}(2003)}]{Moliver2003}%
  \BibitemOpen
  \bibfield  {author} {\bibinfo {author} {\bibfnamefont {S.~S.}\ \bibnamefont
  {Moliver}},\ }\href {\doibase 10.1134/1.1626778} {\bibfield  {journal}
  {\bibinfo  {journal} {Tech. Phys.}\ }\textbf {\bibinfo {volume} {48}},\
  \bibinfo {pages} {1449} (\bibinfo {year} {2003})}\BibitemShut {NoStop}%
\bibitem [{\citenamefont {Goss}\ \emph {et~al.}(1996)\citenamefont {Goss},
  \citenamefont {Jones}, \citenamefont {Breuer}, \citenamefont {Briddon},\ and\
  \citenamefont {\"{O}berg}}]{Goss1996}%
  \BibitemOpen
  \bibfield  {author} {\bibinfo {author} {\bibfnamefont {J.P.}~\bibnamefont
  {Goss}}, \bibinfo {author} {\bibfnamefont {R.}~\bibnamefont {Jones}},
  \bibinfo {author} {\bibfnamefont {S.J.}~\bibnamefont {Breuer}}, \bibinfo
  {author} {\bibfnamefont {P.R.}~\bibnamefont {Briddon}}, \ and\ \bibinfo
  {author} {\bibfnamefont {S.}~\bibnamefont {\"{O}berg}},\ }\href
  {http://www.ncbi.nlm.nih.gov/pubmed/10062116} {\bibfield  {journal} {\bibinfo
   {journal} {Phys. Rev. Lett.}\ }\textbf {\bibinfo {volume} {77}},\ \bibinfo
  {pages} {3041} (\bibinfo {year} {1996})}\BibitemShut {NoStop}%
\bibitem [{\citenamefont {Tinkham}(1964)}]{Tinkham1964}%
  \BibitemOpen
  \bibfield  {author} {\bibinfo {author} {\bibfnamefont {M.}~\bibnamefont
  {Tinkham}},\ }\href@noop {} {\emph {\bibinfo {title} {{Group theory and
  quantum mechanics}}}}\ (\bibinfo  {publisher} {Dover Publications},\ \bibinfo
  {address} {Mineola, N.Y.},\ \bibinfo {year} {1964})\BibitemShut {NoStop}%
\bibitem [{\citenamefont {Walker}(1979)}]{Walker1979}%
  \BibitemOpen
  \bibfield  {author} {\bibinfo {author} {\bibfnamefont {J.}~\bibnamefont
  {Walker}},\ }\href {http://iopscience.iop.org/0034-4885/42/10/001} {\bibfield
   {journal} {\bibinfo  {journal} {Reports Prog. Phys.}\ }\textbf {\bibinfo
  {volume} {42}},\ \bibinfo {pages} {1605} (\bibinfo {year}
  {1979})}\BibitemShut {NoStop}%
\bibitem [{\citenamefont {Goss}\ \emph {et~al.}(2007)\citenamefont {Goss},
  \citenamefont {Briddon},\ and\ \citenamefont {Shaw}}]{Goss2007}%
  \BibitemOpen
  \bibfield  {author} {\bibinfo {author} {\bibfnamefont {J.~P.}\ \bibnamefont
  {Goss}}, \bibinfo {author} {\bibfnamefont {P.~R.}\ \bibnamefont {Briddon}}, \
  and\ \bibinfo {author} {\bibfnamefont {M.~J.}\ \bibnamefont {Shaw}},\
  }\href@noop {} {\bibfield  {journal} {\bibinfo  {journal} {Phys. Rev. B}\
  }\textbf {\bibinfo {volume} {76}},\ \bibinfo {pages} {075204} (\bibinfo {year}
  {2007})}\BibitemShut {NoStop}%
\bibitem [{\citenamefont {Brown}\ and\ \citenamefont {Rand}(1995)}]{Brown1995}%
  \BibitemOpen
  \bibfield  {author} {\bibinfo {author} {\bibfnamefont {S.}~\bibnamefont
  {Brown}}\ and\ \bibinfo {author} {\bibfnamefont {S.}~\bibnamefont {Rand}},\
  }\href@noop {} {\bibfield  {journal} {\bibinfo  {journal} {J. Appl. Phys.}\
  }\textbf {\bibinfo {volume} {78}},\ \bibinfo {pages} {4069} (\bibinfo {year}
  {1995})}\BibitemShut {NoStop}%
\bibitem [{\citenamefont {Neu}\ \emph {et~al.}(2011)\citenamefont {Neu},
  \citenamefont {Fischer}, \citenamefont {Gsell}, \citenamefont {Schreck},\
  and\ \citenamefont {Becher}}]{Neu2011b}%
  \BibitemOpen
  \bibfield  {author} {\bibinfo {author} {\bibfnamefont {E.}~\bibnamefont
  {Neu}}, \bibinfo {author} {\bibfnamefont {M.}~\bibnamefont {Fischer}},
  \bibinfo {author} {\bibfnamefont {S.}~\bibnamefont {Gsell}}, \bibinfo
  {author} {\bibfnamefont {M.}~\bibnamefont {Schreck}}, \ and\ \bibinfo
  {author} {\bibfnamefont {C.}~\bibnamefont {Becher}},\ }\href {\doibase
  10.1103/PhysRevB.84.205211} {\bibfield  {journal} {\bibinfo  {journal} {Phys.
  Rev. B}\ }\textbf {\bibinfo {volume} {84}},\ \bibinfo {pages} {205211}
  (\bibinfo {year} {2011})}\BibitemShut {NoStop}%
\bibitem [{\citenamefont {D'Haenens-Johansson}\ \emph
  {et~al.}(2011)\citenamefont {D'Haenens-Johansson}, \citenamefont {Edmonds},
  \citenamefont {Green}, \citenamefont {Newton}, \citenamefont {Davies},
  \citenamefont {Martineau}, \citenamefont {Khan},\ and\ \citenamefont
  {Twitchen}}]{DHaenens2011}%
  \BibitemOpen
  \bibfield  {author} {\bibinfo {author} {\bibfnamefont {U.~F.~S.}\
  \bibnamefont {D'Haenens-Johansson}}, \bibinfo {author} {\bibfnamefont
  {A.~M.}\ \bibnamefont {Edmonds}}, \bibinfo {author} {\bibfnamefont {B.~L.}\
  \bibnamefont {Green}}, \bibinfo {author} {\bibfnamefont {M.~E.}\ \bibnamefont
  {Newton}}, \bibinfo {author} {\bibfnamefont {G.}~\bibnamefont {Davies}},
  \bibinfo {author} {\bibfnamefont {P.~M.}\ \bibnamefont {Martineau}}, \bibinfo
  {author} {\bibfnamefont {R.~U.~A.}\ \bibnamefont {Khan}}, \ and\ \bibinfo
  {author} {\bibfnamefont {D.~J.}\ \bibnamefont {Twitchen}},\ }\href {\doibase
  10.1103/PhysRevB.84.245208} {\bibfield  {journal} {\bibinfo  {journal} {Phys.
  Rev. B}\ }\textbf {\bibinfo {volume} {84}},\ \bibinfo {pages} {245208}
  (\bibinfo {year} {2011})}\BibitemShut {NoStop}%
\bibitem [{\citenamefont {M\"{u}ller}\ \emph {et~al.}()\citenamefont
  {M\"{u}ller}, \citenamefont {Hepp}, \citenamefont {Pingault}, \citenamefont
  {Neu}, \citenamefont {Gsell}, \citenamefont {Schreck}, \citenamefont
  {Sternschulte}, \citenamefont {Steinm\"{u}ller-Nethl}, \citenamefont
  {Becher},\ and\ \citenamefont {Atat\"{u}re}}]{Mueller2013}%
  \BibitemOpen
  \bibfield  {author} {\bibinfo {author} {\bibfnamefont {T.}~\bibnamefont
  {M\"{u}ller}}, \bibinfo {author} {\bibfnamefont {C.}~\bibnamefont {Hepp}},
  \bibinfo {author} {\bibfnamefont {B.}~\bibnamefont {Pingault}}, \bibinfo
  {author} {\bibfnamefont {E.}~\bibnamefont {Neu}}, \bibinfo {author}
  {\bibfnamefont {S.}~\bibnamefont {Gsell}}, \bibinfo {author} {\bibfnamefont
  {M.}~\bibnamefont {Schreck}}, \bibinfo {author} {\bibfnamefont
  {H.}~\bibnamefont {Sternschulte}}, \bibinfo {author} {\bibfnamefont
  {D.}~\bibnamefont {Steinm\"{u}ller-Nethl}}, \bibinfo {author} {\bibfnamefont
  {C.}~\bibnamefont {Becher}}, \ and\ \bibinfo {author} {\bibfnamefont
  {M.}~\bibnamefont {Atat\"{u}re}},\ }\href@noop {} {\bibinfo  {journal}
  {submitted}\ }\BibitemShut {NoStop}%
\bibitem [{Sup()}]{Supplementary}%
  \BibitemOpen
\bibfield  {journal} {  }\href@noop {} {\enquote {\bibinfo {title} {See
  {S}upplemental {M}aterial at [{URL} will be inserted by publisher] for
  details on samples, experimental setup, theoretical model, polarization curve
  simulation and spin state tomography.}}\ }\BibitemShut {NoStop}%
\bibitem [{\citenamefont {Rogers}\ \emph {et~al.}()\citenamefont {Rogers},
  \citenamefont {Jahnke}, \citenamefont {Doherty}, \citenamefont {Dietrich},
  \citenamefont {Mcguinness}, \citenamefont {Christoph}, \citenamefont
  {Teraji}, \citenamefont {Isoya}, \citenamefont {Manson},\ and\ \citenamefont
  {Jelezko}}]{Rogers2013}%
  \BibitemOpen
  \bibfield  {author} {\bibinfo {author} {\bibfnamefont {L.~J.}\ \bibnamefont
  {Rogers}}, \bibinfo {author} {\bibfnamefont {K.~D.}\ \bibnamefont {Jahnke}},
  \bibinfo {author} {\bibfnamefont {M.~W.}\ \bibnamefont {Doherty}}, \bibinfo
  {author} {\bibfnamefont {A.}~\bibnamefont {Dietrich}}, \bibinfo {author}
  {\bibfnamefont {L.}~\bibnamefont {Mcguinness}}, \bibinfo {author}
  {\bibfnamefont {M.}~\bibnamefont {Christoph}}, \bibinfo {author}
  {\bibfnamefont {T.}~\bibnamefont {Teraji}}, \bibinfo {author} {\bibfnamefont
  {J.}~\bibnamefont {Isoya}}, \bibinfo {author} {\bibfnamefont {N.~B.}\
  \bibnamefont {Manson}}, \ and\ \bibinfo {author} {\bibfnamefont
  {F.}~\bibnamefont {Jelezko}},\ }\href@noop {} {\bibinfo  {journal} {in
  preparation}\ }\BibitemShut {NoStop}%
\bibitem [{\citenamefont {Gali}\ and\ \citenamefont {Maze}()}]{Gali2013}%
  \BibitemOpen
\bibfield  {journal} {  }\bibfield  {author} {\bibinfo {author} {\bibfnamefont
  {A.}~\bibnamefont {Gali}}\ and\ \bibinfo {author} {\bibfnamefont {J.~R.}\
  \bibnamefont {Maze}},\ }\href@noop {} {\ }\Eprint
  {http://arxiv.org/abs/1310.2137v1} {arXiv:1310.2137v1} \BibitemShut {NoStop}%
\bibitem [{\citenamefont {Abtew}\ \emph {et~al.}(2011)\citenamefont {Abtew},
  \citenamefont {Sun}, \citenamefont {Shih}, \citenamefont {Dev}, \citenamefont
  {Zhang},\ and\ \citenamefont {Zhang}}]{Abtew2011}%
  \BibitemOpen
  \bibfield  {author} {\bibinfo {author} {\bibfnamefont {T.~A.}\ \bibnamefont
  {Abtew}}, \bibinfo {author} {\bibfnamefont {Y.~Y.}\ \bibnamefont {Sun}},
  \bibinfo {author} {\bibfnamefont {B.-C.}\ \bibnamefont {Shih}}, \bibinfo
  {author} {\bibfnamefont {P.}~\bibnamefont {Dev}}, \bibinfo {author}
  {\bibfnamefont {S.~B.}\ \bibnamefont {Zhang}}, \ and\ \bibinfo {author}
  {\bibfnamefont {P.}~\bibnamefont {Zhang}},\ }\href {\doibase
  10.1103/PhysRevLett.107.146403} {\bibfield  {journal} {\bibinfo  {journal}
  {Phys. Rev. Lett.}\ }\textbf {\bibinfo {volume} {107}},\ \bibinfo {pages}
  {146403} (\bibinfo {year} {2011})}\BibitemShut {NoStop}%
\bibitem [{\citenamefont {Rorison}\ and\ \citenamefont
  {O'Brien}(1984)}]{Rorison1984}%
  \BibitemOpen
  \bibfield  {author} {\bibinfo {author} {\bibfnamefont {J.}~\bibnamefont
  {Rorison}}\ and\ \bibinfo {author} {\bibfnamefont {M.}~\bibnamefont
  {O'Brien}},\ }\href {\doibase 10.1088/0022-3719/17/19/019} {\bibfield
  {journal} {\bibinfo  {journal} {J. Phys. C Solid State Phys.}\ }\textbf
  {\bibinfo {volume} {17}},\ \bibinfo {pages} {3449} (\bibinfo {year}
  {1984})}\BibitemShut {NoStop}%
\bibitem [{\citenamefont {Longuet-Higgins}\ \emph {et~al.}(1958)\citenamefont
  {Longuet-Higgins}, \citenamefont {Opik}, \citenamefont {Pryce},\ and\
  \citenamefont {Sack}}]{LonguetHiggins1958}%
  \BibitemOpen
  \bibfield  {author} {\bibinfo {author} {\bibfnamefont {H.~C.}\ \bibnamefont
  {Longuet-Higgins}}, \bibinfo {author} {\bibfnamefont {U.}~\bibnamefont
  {Opik}}, \bibinfo {author} {\bibfnamefont {M.~H.~L.}\ \bibnamefont {Pryce}},
  \ and\ \bibinfo {author} {\bibfnamefont {R.~A.}\ \bibnamefont {Sack}},\
  }\href {\doibase 10.1098/rspa.1958.0022} {\bibfield  {journal} {\bibinfo
  {journal} {Proc. R. Soc. A Math. Phys. Eng. Sci.}\ }\textbf {\bibinfo
  {volume} {244}},\ \bibinfo {pages} {1} (\bibinfo {year} {1958})}\BibitemShut
  {NoStop}%
\bibitem [{\citenamefont {Ham}(1965)}]{Ham1965}%
  \BibitemOpen
  \bibfield  {author} {\bibinfo {author} {\bibfnamefont {F.~S.}\ \bibnamefont
  {Ham}},\ }\href@noop {} {\bibfield  {journal} {\bibinfo  {journal} {Phys.
  Rev.}\ }\textbf {\bibinfo {volume} {138}},\ \bibinfo {pages} {1727} (\bibinfo
  {year} {1965})}\BibitemShut {NoStop}%
\bibitem [{\citenamefont {Tosatti}\ \emph {et~al.}(1996)\citenamefont
  {Tosatti}, \citenamefont {Manini},\ and\ \citenamefont
  {Gunnarsson}}]{Tosatti1996}%
  \BibitemOpen
  \bibfield  {author} {\bibinfo {author} {\bibfnamefont {E.}~\bibnamefont
  {Tosatti}}, \bibinfo {author} {\bibfnamefont {N.}~\bibnamefont {Manini}}, \
  and\ \bibinfo {author} {\bibfnamefont {O.}~\bibnamefont {Gunnarsson}},\
  }\href@noop {} {\bibfield  {journal} {\bibinfo  {journal} {Phys. Rev. B}\
  }\textbf {\bibinfo {volume} {54}},\ \bibinfo {pages} {17184} (\bibinfo {year}
  {1996})}\BibitemShut {NoStop}%
\bibitem [{\citenamefont {Neu}\ \emph {et~al.}(2012)\citenamefont {Neu},
  \citenamefont {Agio},\ and\ \citenamefont {Becher}}]{Neu2012b}%
  \BibitemOpen
  \bibfield  {author} {\bibinfo {author} {\bibfnamefont {E.}~\bibnamefont
  {Neu}}, \bibinfo {author} {\bibfnamefont {M.}~\bibnamefont {Agio}}, \ and\
  \bibinfo {author} {\bibfnamefont {C.}~\bibnamefont {Becher}},\ }\href@noop {}
  {\bibfield  {journal} {\bibinfo  {journal} {Opt. Express}\ }\textbf {\bibinfo
  {volume} {20}},\ \bibinfo {pages} {19956} (\bibinfo {year}
  {2012})}\BibitemShut {NoStop}%
\end{thebibliography}

%

\end{document}